\documentstyle[sprocl]{article}
\def\be{\begin{equation}}
\def\ee{\end{equation}}
\def\bea{\begin{eqnarray}}
\def\eea{\end{eqnarray}}
\begin{document}

\title{FINAL STATE INTERACTIONS IN BOSE-EINSTEIN CORRELATIONS
\footnote{To be published in proceedings of 7th International Workshop 
on Multiparticle production ``Correlation and Fluctuations'' 
(Nijmegen, June 30-July 6, 1996); World Scientific. }}

\author{M.~BIYAJIMA$^1$,~T.~MIZOGUCHI$^2$,~
T.~OSADA$^{1,3}$ and G.~WILK$^{1,4}$}

\address{$^1$Department of Physics, Faculty of Science,
Shinshu University,\\ Matsumoto 390, Japan\\
$^2$Toba National College of Maritime Technology, Toba 517, Japan\\
$^3$Department of Physics, Tohoku University, Sendai 980, Japan\\
$^4$Soltan Institute for Nuclear Studies, Nuclear Theory Department\\
\it Ho\.za 69, PL-00-681 Warsaw, Poland}

\maketitle\abstracts{
We are presenting here the new formulae for Bose-Einstein
correlations (BEC) which contain effects of final state interactions (FSI)
of both strong (in $s$-wave) and electromagnetic origin. 
We demonstrate the importance of FSI in BEC by analysing  
data for $e^+e^-$ annihilation and for heavy collisions.
The inclusion of FSI results in the practical elimination 
(at least in $e^+e^-$ data) of the so called
degree of coherence parameter $\lambda$ (which becomes equal unity) 
and the long range parameter $\gamma$ (which is now equal zero).
}
\section{Introduction}
Bose-Einstein correlations (BEC) are the one of  the most
important current topics in high energy collisions (in particular in
heavy-ion reactions). One of the 
interesting problems present in the BEC is the physical meaning 
of the degree of coherence parameter $\lambda$ and the
long range correlation parameter $\gamma$ which are usually 
introduced by hand when analysing BEC data by means of the 
so-called standard formula~\cite{biya83}:
\begin{equation}
  N^{(\pm\pm)}/N^{BG} = c\left[1 + \lambda E_{2B} \right]
  (1 + \gamma Q).
\end{equation}
The function $E_{2B}$ appearing here is called the exchange function
and $c$ is the so-called normalization parameter (also introduced by
hand but we shall not discuss it here).

Our approach to this problem is based on the observation that
BEC formula Eq.(1), which has been obtained by using the plane wave
approximation (asymptotic states) for both observed particles, 
should be corrected for the effect of final state 
interactions (FSI) which can be of strong and electromagnetic type (Coulomb
interactions). 
We have recently obtained several theoretical 
formulae for the BEC including effects of FSI  of the strong type
(in the $s$-wave, isospin $I=2$ channel) \cite{biya95zpc}, of 
the Coulomb interactions \cite{biya95plb,biya96plb} and of both types of FSI acting together
\cite{osa96zpc}.

The second order BEC without FSI is usually presented as the following
convolution 
\begin{eqnarray}
N^{(\pm \pm)}/N^{BG} = \int_0^\infty \rho(x_1)\rho(x_2) 
             \vert A_{12} \vert^2 d^3x_1d^3x_2, \nonumber\\
                    = \int_0^\infty \rho(R)\rho(r) 
             \vert A_{12} \vert^2 d^3Rd^3r
\end{eqnarray}
of the single particle source density functions $\rho$ and the squared
two particle amplitude 
\begin{eqnarray}
        A_{12} = \frac{1}{\sqrt{2}} [\exp\{(ip_1(x_A-x_1) + 
                 ip_2(x_B-x_2)\} \nonumber\\
               + \exp\{(ip_1(x_A-x_2) +ip_2(x_B-x_1)\}]  
\end{eqnarray}
which is symmetrized accordingly, cf. Fig. 1.
\vspace*{4.5cm}
\begin{center}
{\small Fig. 1 Identical boson exchange diagram.}
\end{center}
Assuming now the following Gaussian distribution for the source function, 
$  \rho(r) = e^{-r^2/2\beta^2} $, 
one obtains Eq. (1) with $ E_{2B}= e^{-\beta^2 Q^2/2}$ (and with $c=1$,
$\lambda =1$ and $\gamma=0$). 
 
In the next section we shall present how to correct this formula
for the presence of strong interactions FSI 
which are seen in the phase shift analysis of
the $\pi \pi$ and  $K_s^0 K_s^0$ correlations.
The inclusion of Coulomb interactions, which are important when produced
bosons are charged, demands however a profound change of the two particle 
amplitude $A_{12}$. This is discussed in Section 3. 
It is shown there also that, we can obtain information 
on the interaction range from data for unlike charged pairs
$\pi^+ \pi^-$.  Section 4 deals with the most general case when strong
and Coulomb FSI coexist together.
Our concluding remarks are presented in the last section.

\section{Final state interaction in BEC for neutral particles}

FSI of the strong type are limited (due to the short range
of strong interactions) to a small number of partial waves (in practice
to $s$-wave only) and it is therefore sufficient to use 
the following expression for the amplitude describing the system of
two identical bososns in their rest frame 
\cite{biya95zpc,bowler88}:
\begin{equation}
  A_{12} =
  \frac{1}{\sqrt{2}}e^{i(\mbox{\small {\bf p}}_1 + 
  \mbox{\small {\bf p}}_2) \cdot \mbox{\small {\bf R}}}
  \left[\frac{\left(e^{2i\delta} - 1\right)}{ikr}e^{ikr}
  + e^{i\mbox{\small {\bf k}}\cdot \mbox{\small {\bf r}}}
  + e^{-i\mbox{\small {\bf k}}\cdot \mbox{\small {\bf r}}}\right],
  \label{eq:a3}
\end{equation}
Here ${\bf R} = ({\bf r}_1 + {\bf r}_2)/2$, ${\bf k} = 
({\bf p}_1 - {\bf p}_2)/2$, and ${\bf r} = ({\bf r}_1 - 
{\bf r}_2)$
and $\delta$ 
denotes the phase shift describing the corresponding FSI. 
The data of phase shifts of $ \pi \pi$ \cite{suzuki87} and $K_s^0K_s^0$ 
\cite{balo95pan}, which were used in our analysis, are shown in Fig. 2. 
\vspace*{5cm}\\
\begin{center}
{\small  Fig. 2. Phase shifts for $ \pi \pi$  and $K_s^0K_s^0$
collisions.~(cf. [8],[9] for details.)}
\end{center}
After some algebra 
we obtain a following new formula of the BEC containing the strong FSI
given by the respective phase shift function $\delta$:,
\begin{eqnarray}
  &&N^{(2-)}/N^{BG} = 1 + \lambda \left\{e^{-\beta^2Q^2/2}\mbox{Re}
  [\mbox{erfc}(z)]
  + \frac{8\sin^2 \delta}{\beta^2Q^2}e^{-\beta^2Q^2/2}\mbox{Re}
  [\mbox{erfc}(z)]\right.\nonumber \\
  &&\hspace{4cm}\left.+ \frac{8\sin \delta \cos \delta}{\beta^2Q^2}
  e^{-\beta^2Q^2/2}\mbox{Im}[\mbox{erfc}(z)] \right\},
  \label{eq:a9}
\end{eqnarray}
where $z = - i\beta Q/\sqrt{2}$,
and where 
$  \mbox{Re}[\mbox{erfc}(- i\beta Q/\sqrt{2})] = 1$
relation was used 
(the degree of coherence parameter $\lambda$ was added by hand here).
Notice that for $\delta \to 0$, i.e., when the FSI is switched off, 
we are recovering standard formula given by Eq. (1) with $E =
e^{-\beta^2Q^2/2}$ (if we add by hand additional parameters
$c$ and $\gamma$ in the form of long range correlation factor $(1 +
\gamma Q)$).
In Fig. 3 we present our analysis of BEC data for $K_s^0K_s^0$ pair 
production using the above approach \cite{delphi96}.
\vspace*{5cm}
\begin{center}
{\small  Fig. 3. Analyses of the data of $K_s^0K_s^0$ pair.}\\
\end{center} 
\section{Final state interaction in BEC for charged particles - Coulomb
interactions}
In the case of Coulomb type of FSI 
the amplitude $A_{12}$ has to be described by the so called Coulomb 
wave functions: 
\begin{eqnarray}
 A_{12} &=& \frac 1{\sqrt{2}} [ \Psi({\bf k},{\bf r}) +
\Psi_S({\bf k},{\bf r}) ]\:,\\
 \Psi({\bf k},{\bf r}) &=& \Gamma(1+i\eta)e^{-\pi \eta/2}
e^{i{\bf k}\cdot {\bf r}}
\Phi(-i\eta;1;ikr(1 - \cos \theta))\:, \nonumber\\
 \Psi_S({\bf k},{\bf r}) &=& \Gamma(1+i\eta)e^{-\pi \eta/2}
e^{-i{\bf k}\cdot {\bf r}}
\Phi(-i\eta;1;ikr(1 + \cos \theta))\:,\nonumber
\end{eqnarray}
where $r = x_1 - x_2$, the parameter 
$\eta = m_{\mbox{\scriptsize red}} \alpha/k$,  
the momentum transfer $Q = (p_1 - p_2) = 2k$ and $\Phi$ denotes the 
confluent hypergeometric function $\Phi$ \cite{schiff}.
The BEC formula now reads:
\begin{eqnarray}
N^{(\pm \pm)}/ N^{\mbox{\scriptsize BG}} &=& \frac 1{G(2k)} \int \rho(R)
                  d^3R \int \rho(r) d^3r |A_{12}|^2 \nonumber\\
 &=& \sum_{n=0}^{\infty} \sum_{m=0}^{\infty} \frac{(-i)^n(i)^m}
     {n+m+1} (2k)^{n+m}I_R(n,m) A_n A_m^* \nonumber\\
 &&\times\left[ 1+ \frac{n!m!}{(n+m)!}
  \left( 1+\frac{n}{i\eta}\right) \left( 1-\frac{m}{i\eta}\right)
  \right ] \label{eq:BEC}\\
  &=& (1 + \Delta_{\mbox{\scriptsize 1C}}) + (\Delta_{\mbox{\scriptsize EC}} 
       + E_{\mbox{\scriptsize 2B}}),\label{eq:result}
\end{eqnarray}
where $G(2k) = 2\pi \eta /(e^{2\pi \eta} - 1)$ denotes Gamow factor,
the first and the second parentheses in Eq.~(\ref{eq:result})
correspond to the first and the second terms in
Eq.~(\ref{eq:BEC}) and (cf. ref. [3]) 
$$
I_R(n,m) = 4 \pi \int dr\, r^{2 + n + m} \rho (r),\qquad
A_n = \frac{\Gamma(i\eta + n)}{\Gamma(i\eta)}\frac{1}{(n!)^2}.
$$
This formula, as it was shown in ref.[3],  
can be also used to analyse data on unlike sign  
$\pi^+\pi^-$ pair production 
\cite{afan90}. This data cover 
the Coulomb interaction region at small momentum
transfer $ Q \le $ 40 MeV/c in $ P $ + Ta $\to \pi^+  + \pi^- $ + X
at proton energy 70 GeV and have been so far analysed by using 
only Gamow factor correction\cite{afan90}.
However, when analysed by using instead our Coulomb formula with 
$\eta=m_{\pi}\alpha/2k$, 
\begin{equation}
     N^{(+-)}/N^{BG} = G(-\eta)(1 + \Delta_{1C}(-\eta))
\end{equation}
they can provide us information on the interaction region in
$p$ + Ta $\to  \pi^+ + \pi^- $ + X  collisions which turns out
to be about 2 fm, cf. Fig. 4.
It should be noticed that the interaction range cannot be estimated
from the analysis performed only by using the Gamow factor.
\vspace*{5cm}\\
\begin{center}
{\small Fig. 4.   Analysis of data for the $\pi^+\pi^-$ pair production. } 
\end{center}
When one applies our formulae to single pion production in the external
Coulomb field one founds an apparent quasi-scaling behavior \cite{ratio96}
in the pion production 
yield $N^{\pi}(\beta, Z_{eff}) = 1 + \Delta_{1C}$, namely one
observes that
\begin{equation}
N^{\pi}(Q;~\beta, Z_{eff}) \approx  N^{\pi}(Q;~\lambda \times \beta, 
\lambda \times Z_{eff}),\quad (1 \le \lambda \le 3).
\end{equation}
Because of this property, we have found that 
it is difficult to estimate the magnitude  of 
nuclear fragments ($Z_{eff}$) produced in the 
central region in heavy-ion collisions 
from the yield ratio $N^{\pi^{+}}/N^{\pi^{-}}$ only. 

\section{ Final state interaction including both 
strong and Coulomb interactions}
To describe a pair of the identical bosons including both strong and Coulomb 
interactions, we have to
symmetrize the total wave function in the following way:
\begin{equation}
  A_{12} = \frac{1}{\sqrt{2}}
 [ \Psi_{\mbox{\scriptsize C}}({\bf k},{\bf r})
+ \Psi_{\mbox{\scriptsize C}}^{S}({\bf k},{\bf r})
 + \Phi_{\mbox{\scriptsize st}}({\bf k},{\bf r})
+ \Phi_{\mbox{\scriptsize st}}^{S}({\bf k},{\bf r})].
\end{equation}
Here $\Psi_{\mbox{\scriptsize C}}({\bf k},{\bf r})$ denotes Coulomb wave
function described above, superscript $S$ denotes the symmetrization 
of the wave
function and function $ \Phi_{\mbox{\scriptsize st}}({\bf k},{\bf r}) $ 
stands for the wave function induced by strong interactions. Assuming a
source function $\rho(r)$ we obtain the following expression for the
BEC in this case: 
\begin{eqnarray}
   \frac{N^{(\pm\pm)}}{N^{BG}}&=& \frac{1}{G(2k)} \int_{}^{}
   \rho(r) d^3r|A_{12}|^2,
                  \nonumber \\
   &=& I_{\mbox{\scriptsize C}} + I_{\mbox{\scriptsize Cst}} + 
   I_{\mbox{\scriptsize st}}, \\
   I_{\mbox{\scriptsize C}}   &=& 
         (1 + \Delta_{\mbox{\scriptsize 1C}}) + (E_{\mbox{\scriptsize 2B}} +
\Delta_{\mbox{\scriptsize EC}}), \nonumber\\
   I_{\mbox{\scriptsize  Cst}} &=& 2\Re
               \left[
        \frac{2}{k} (2k)^{i\eta}
   \exp{(-i(\eta_0+\delta_{0}^{(2)}))} \sin\delta_{0}^{(2)}
    \sum_{n=0}^{\infty} I_{\mbox{\scriptsize R2}}(1+n) A_2(n,0)
               \right], \nonumber\\
    I_{\mbox{\scriptsize  st}}  &=& \frac{2}{k^2}
I_{\mbox{\scriptsize R1}}(0) \sin^2\delta_{0}^{(2)}. \nonumber
\end{eqnarray} 
Explicit expressions for quantities present here are given in
ref.[5]. 
It is also shown there that our method is equivalent to the numerical 
solution of the Sch\"odinger equation with strong and Coulomb potentials
\cite{pcz90}. 
Our final formula (containing three additional parameters:
$c$, $\lambda$ and $\gamma$, added by hand) is therefore given
as:
\begin{eqnarray}
N^{(\pm\pm)}/N^{\mbox{\scriptsize BG}}
(Q = 2k) &=& c~( 1 + \Delta_{\mbox{\scriptsize 1C}}
  + \Delta_{\mbox{\scriptsize EC}}+ I_{\mbox{\scriptsize Cst}} + 
I_{\mbox{\scriptsize st}}) \nonumber \\
& &\hspace*{-2cm} \times
      \left[
           1 +
     \lambda \frac{ E_{\mbox{\scriptsize 2B}} }
    { 1 + \Delta_{\mbox{\scriptsize 1C}}+\Delta_{\mbox{\scriptsize EC}}+
I_{\mbox{\scriptsize Cst}} + I_{\mbox{\scriptsize st}} }
      \right]
      (1 + \gamma Q).
      \hspace{1cm} \label{eq:ratio}
\end{eqnarray}
It should be noted that the normalization $c$ and an effective
degree of coherence, i.e., the denominator of the ratio
$E_{\mbox{\scriptsize 2B}}/( 1 + \Delta_{\mbox{\scriptsize 1C}} 
+ \Delta_{\mbox{\scriptsize EC}} +
I_{\mbox{\scriptsize Cst}} + I_{\mbox{\scriptsize st}})$, 
are related to each other.
For the sake of reference we shall also use in our analyses the
standard formula as given be Eq. (1) (with $E_{2B} =
\exp(-\beta^2Q^2/2)$).
We apply now our formulae to data for $e^+e^-$ annihilation\cite{aihara85,opal91}. Results of our analyses 
 are shown in Fig. 5 and Table I.
As seen in Table I, our (Eq.(13)) estimated values of the degree of coherence
parameter $\lambda$ are systematically larger (approaching unity)
than those obtained by the standard formula (Eq.(1)).
Similarly, the long range correlation
parameter $\gamma$ approaches now (approximately) zero. 
\vspace*{5cm}
\begin{center}
{\small Fig. 5. Example of the analysis of data in $ e^+e^-$ 
annihilation for TPC \cite{aihara85} and \\OPAL \cite{opal91} 
collaborations.}
\end{center}

\begin{table}[h]
\begin{center}
\begin{tabular}{cccccc}
\hline
             & $\beta$~[fm]      & $\lambda$
              & $\gamma$         & $c$             & $\chi^2/$~NDF \\
\hline\hline
TPC           &&&&& \\
ref.[14] & $0.92 \pm0.06^\ast$  & $0.61\pm0.05$
        &  $-$~~~~~~~       & $-$~~~~~~~      & $-$~~~~       \\
Eq.(13) & $0.74\pm0.05$      & $1.10\pm0.04$
        &  $-0.00\pm0.02$ & $1.00\pm0.02$ & $44.2/35$     \\
Eq.(1) & $0.91\pm0.06$      & $0.61\pm0.05$
        &  $0.08\pm0.03$  & $0.88\pm0.02$ & $41.0/35$     \\
\hline
OPAL          &&&&& \\
ref.[15] & $1.12\pm0.02^\ast$ & $0.85\pm0.03$
        &  $-$~~~~~~~       & $-$~~~~~~~      & $336/73$      \\
Eq.(13) & $1.09\pm0.04$      & $1.04\pm0.03$
        &  $0.00\pm0.00$  & $0.99\pm0.01$ & $124.4/74$    \\
Eq.(1) & $1.34\pm0.04$      & $0.71\pm0.04$
        &  $0.04\pm0.00$  & $0.94\pm0.00$ & $118.7/74$    \\
\hline
\end{tabular}
\caption{Estimated parameters}
\end{center}
\end{table}

\section{Concluding remarks}
We have presented several analytic formulae for BEC including the Coulombic
and strong FSI obtained by us recently. 
Combining the seamless fitting method \cite{biya95plb} and the CERN
MINUIT program in Eq.(13) we have analysed different sets of BEC data
showing respectively:
\begin{itemize}
\item[$(1)$] The role of strong FSI by using data on  
$K_s^0K_s^0$ phase shifts;
\item[$(2)$] The ability to obtain by using our method the range of
interaction from precise data on $\pi^+\pi^-$ production;
\item[$(3)$] The possibility that $\lambda < 1$ and $\gamma \neq 0$
values of parameters encountered in the standard analyses of data could
be a reflection of the combined action of strong and Coulombic FSI which
were not taken properly into account there.
In fact, also the values of the source size parameter reported by
various collaborations (after using the relation: $\beta =
\sqrt{2}R$) and obtained by the standard formula Eq.(1) are
systematically larger than values estimated by our formula Eq.(13).
\end{itemize}

\noindent 
{\bf Acknowledgments:}~~~
One of authors (M. B.) is partially indebted to the Yamada 
Science Foundation for financial support for his traveling expenses.
G.W. thanks also Yamada Science Foundation for supporting his stay at
Shinshu University, Matsumoto. 
The work was partially supported by the Grant-in Aid for Scientific 
Research from the Ministry of Education, Science and Culture 
(No. 06640383) and (No. 08304024). We are indebted to A. Tomaradze for 
his kind sending data.\\

\noindent 
{\bf References}\\

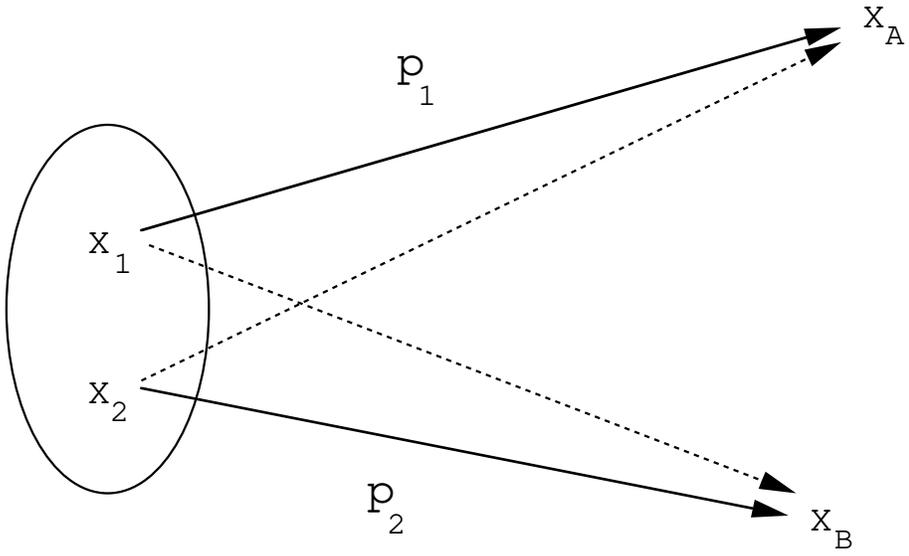
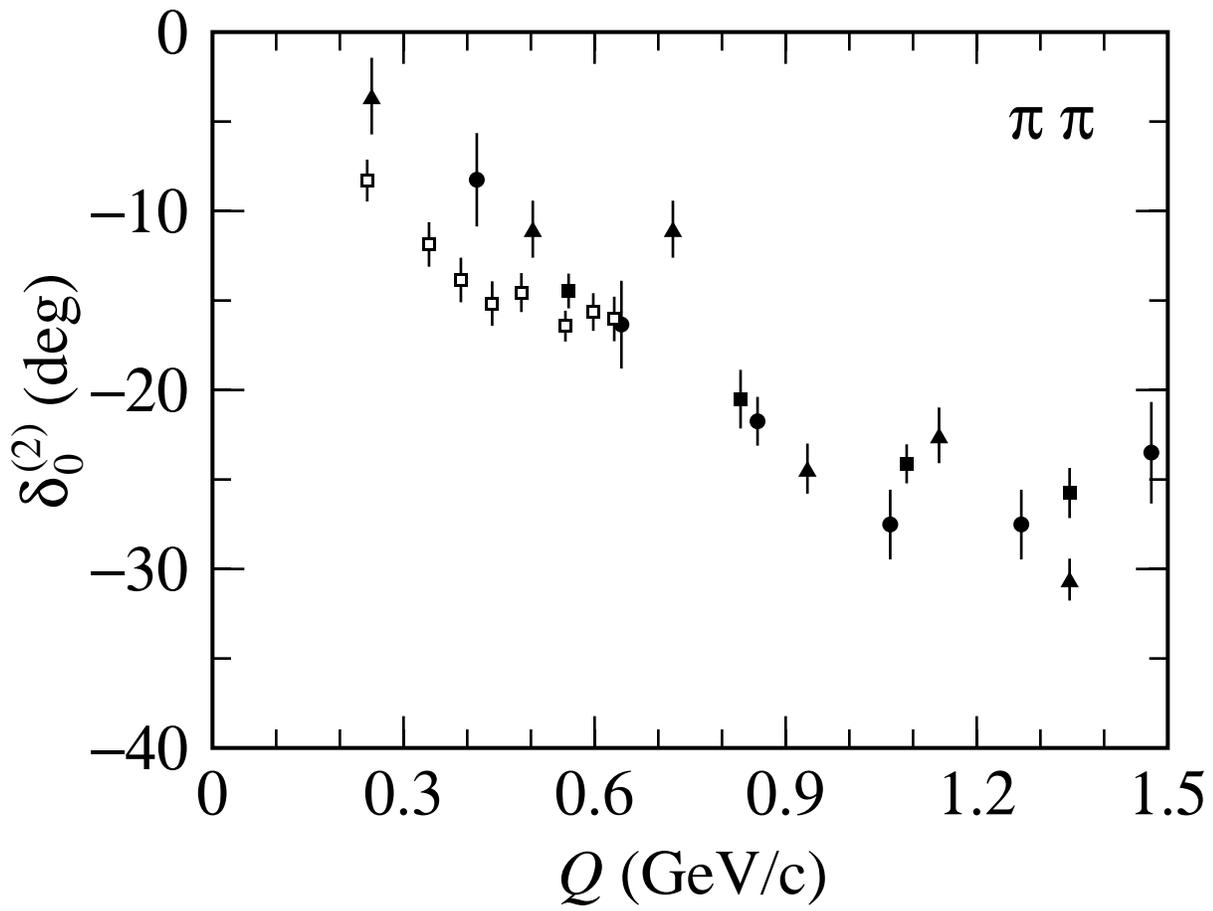
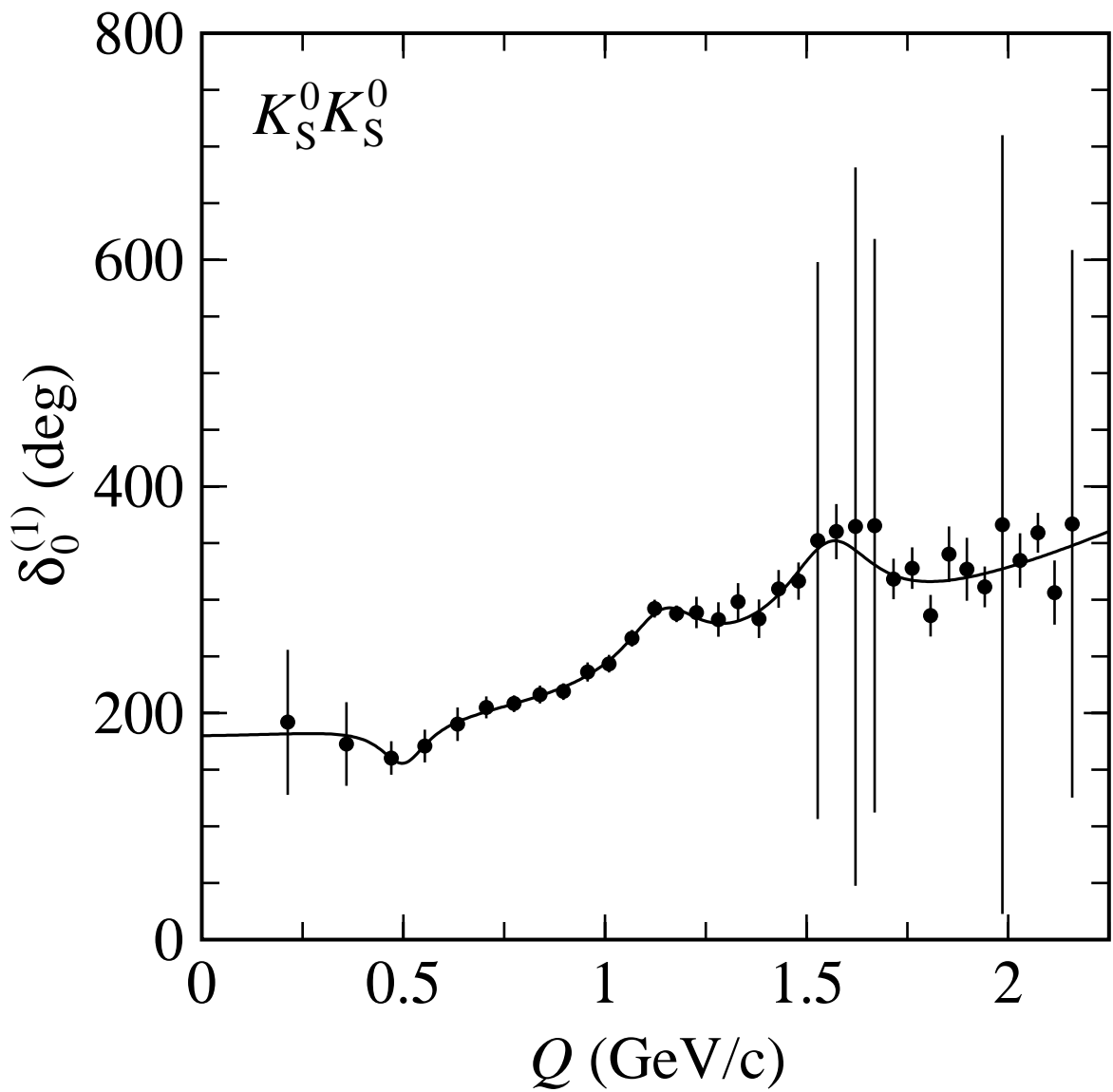
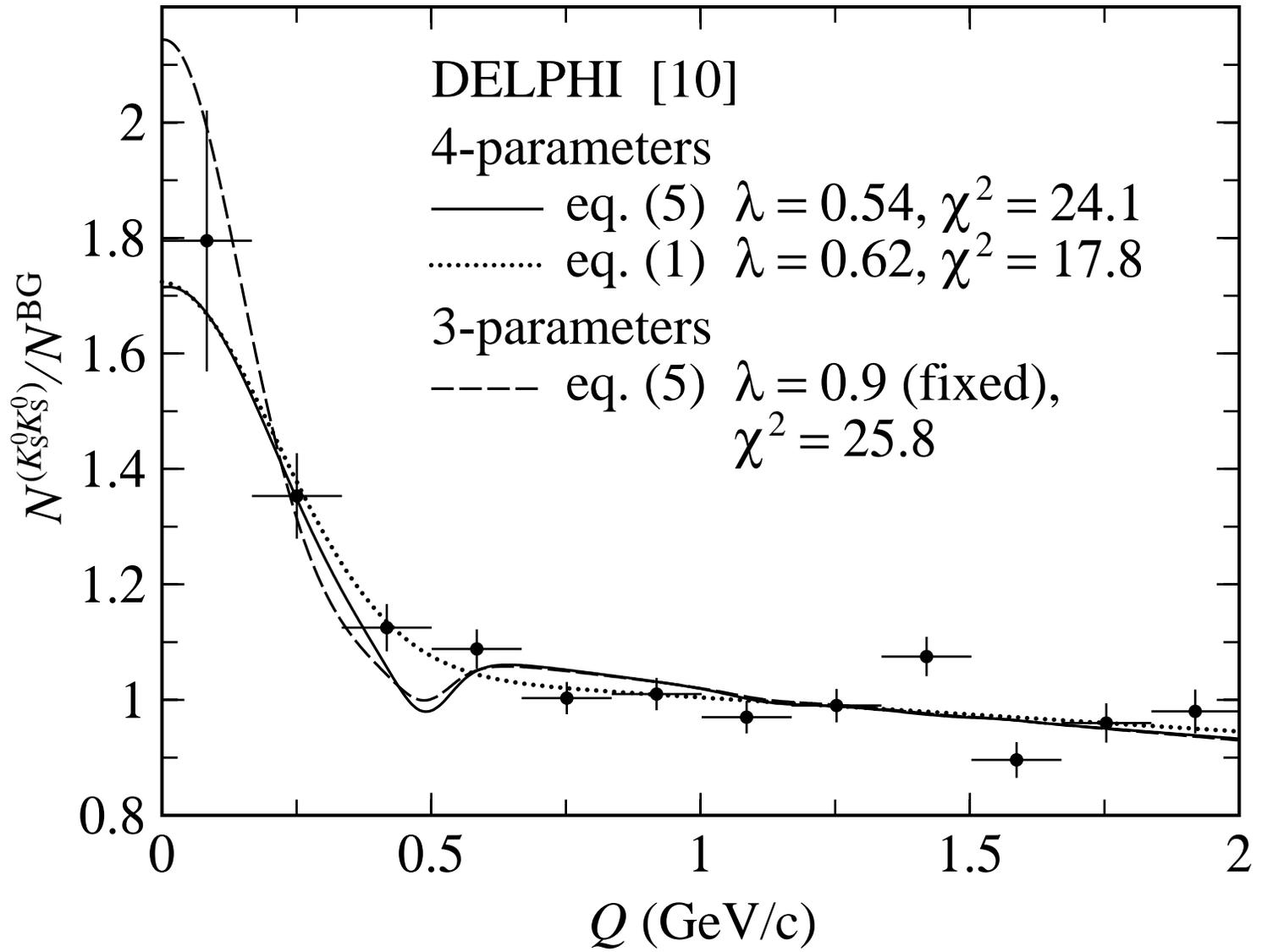
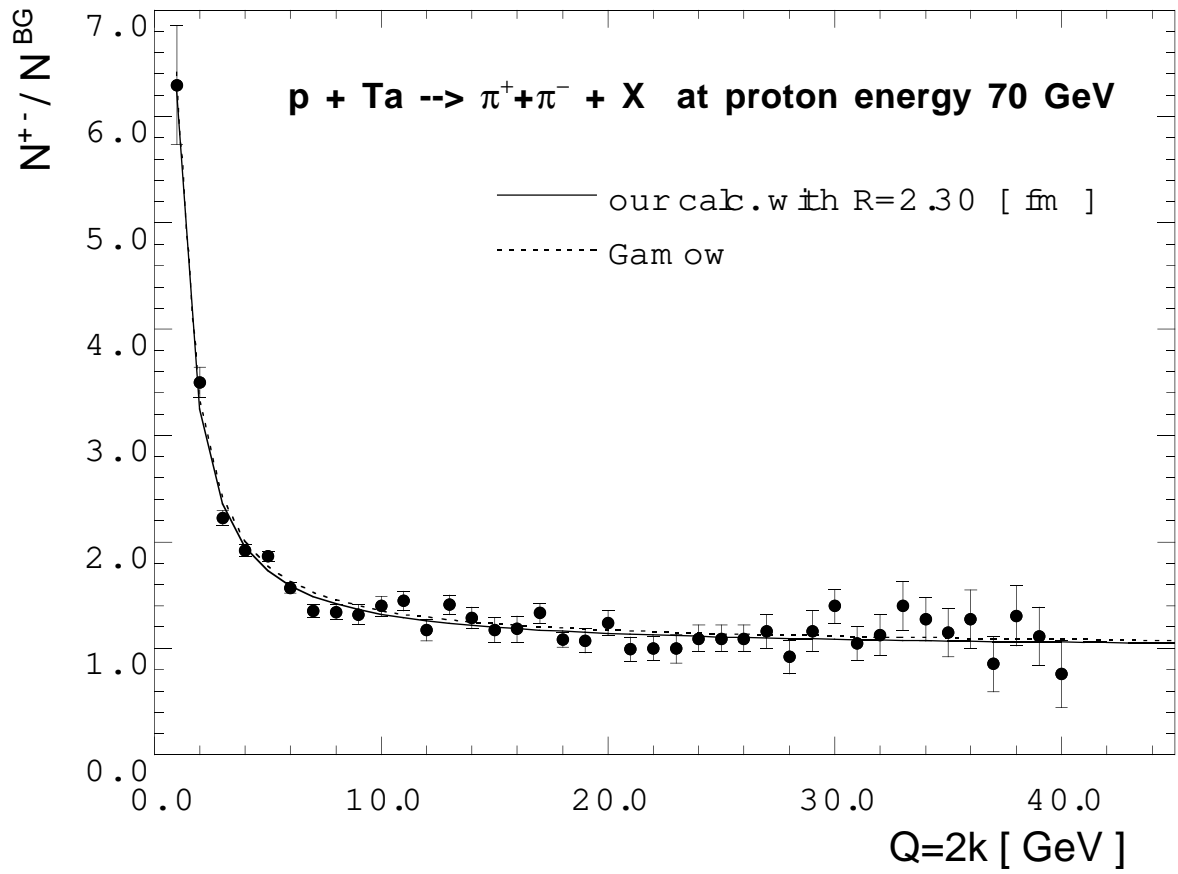
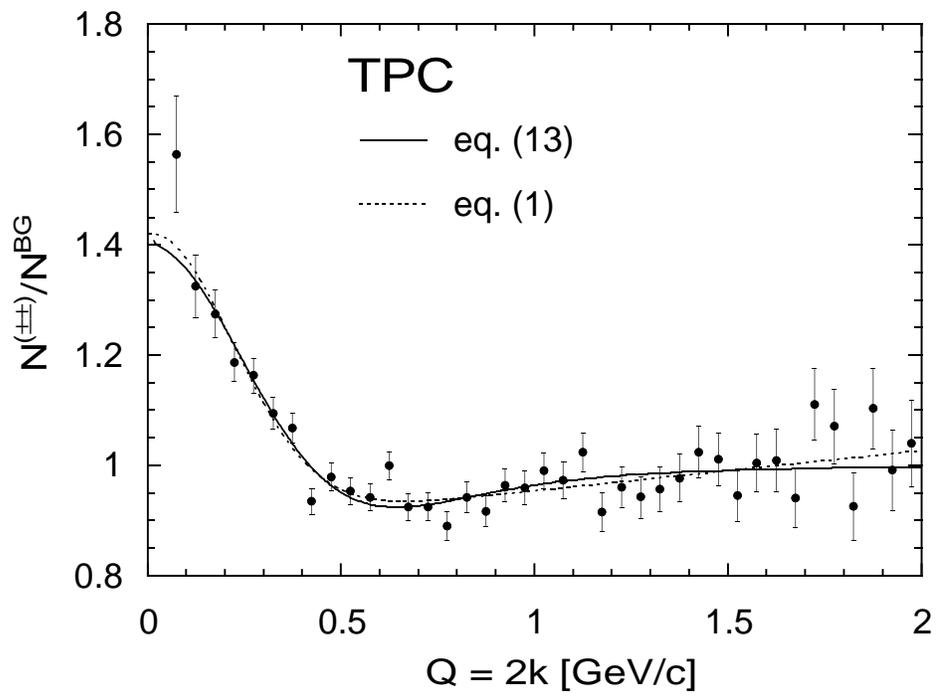
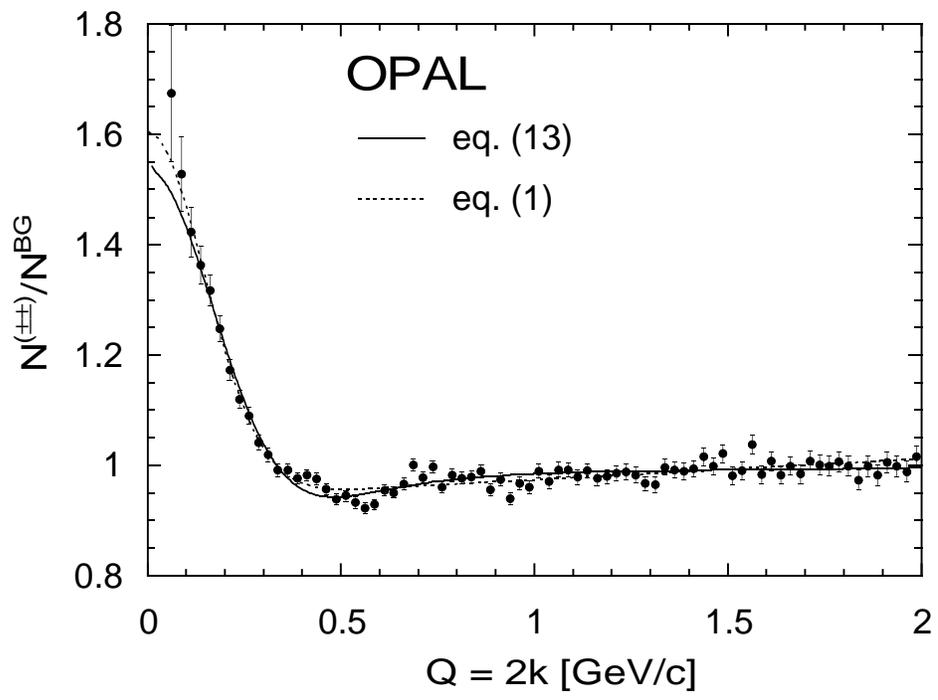
\end{document}